\documentclass[twocolumn,aps,showpacs,preprintnumbers,superscriptaddress,floatfix]{revtex4}

\usepackage[centertags,leqno,sumlimits,intlimits,namelimits]{amsmath}
\usepackage{amsfonts,amsbsy,amsxtra}
\usepackage{amssymb,latexsym}
\usepackage{bm}
\usepackage{ifpdf}
\ifpdf
  \usepackage[pdftex]{graphicx}
  \usepackage[pdftex]{color}
\else
  \usepackage[dvips]{graphicx}
  \usepackage[dvips]{color}
\fi
\usepackage{textcomp}
\usepackage{ifthen}
\usepackage[utf8]{inputenc}
\ifpdf
  \usepackage[
    unicode=true,
    pdftex=true,
    plainpages=false,
    bookmarks=true,
    bookmarksnumbered=true,
    bookmarksopen=true,
    bookmarksopenlevel=2,
    colorlinks=true,
    anchorcolor=black,
    linkcolor=black,
    urlcolor=black,
    citecolor=black,
    hyperindex=true
  ]{hyperref}
\else
  \usepackage[
    unicode=true,
    dvips=true,
    plainpages=false,
    bookmarks=true,
    bookmarksnumbered=true,
    bookmarksopen=true,
    bookmarksopenlevel=2,
    colorlinks=true,
    anchorcolor=black,
    linkcolor=black,
    urlcolor=black,
    citecolor=black,
    hyperindex=true
  ]{hyperref}
\fi

\usepackage{color}
\usepackage{soul}
\sethlcolor{green}
\setstcolor{red}
\newcommand{\antibar}[1]{%
  \overline{#1}%
}
\newcommand{\PPh}{\ensuremath{\gamma}}
\newcommand{\PvPh}{\ensuremath{\gamma^*}}
\newcommand{\PPom}{\ensuremath{\mathbb{P}}}
\renewcommand{\Pr}{\ensuremath{\rho}}
\newcommand{\Prz}{\ensuremath{\rho^0}}
\newcommand{\Pf}{\ensuremath{f_0}}
\newcommand{\PJ}{\ensuremath{J\!/\!\psi}}
\newcommand{\Po}{\ensuremath{\omega}}
\newcommand{\Ppi}{\ensuremath{\pi}}

\newcommand{\Ppip}{\ensuremath{\pi^+}}
\newcommand{\Ppim}{\ensuremath{\pi^-}}

\newcommand{\Pep}{\ensuremath{e^+}}
\newcommand{\Pem}{\ensuremath{e^-}}

\newcommand{\Pqq}{\ensuremath{q}}
\newcommand{\Pqqbar}{\ensuremath{\antibar{q}}}
\newcommand{\Pp}{\ensuremath{p}}
\newcommand{\Pn}{\ensuremath{n}}

\newcommand{\PA}{\ensuremath{A}}
\newcommand{\PAu}{\ensuremath{\text{Au}}}

\newcommand{\twopion}{\ensuremath{\Ppip\Ppim}}
\newcommand{\fourpion}{\ensuremath{\Ppip\Ppim\Ppip\Ppim}}

\newcommand{\lrbrk}[1]{{\left({#1}\right)}}
\newcommand{\lrBrk}[1]{{\left[{#1}\right]}}
\newcommand{\lrabs}[1]{{\left|{#1}\right|}}
\newcommand{\abs}[1]{{|{#1}|}}

\newcommand{\D}[2][]{\operatorname{d^{#1} \mathnormal{#2}}}

\newcommand{\pT}{\ensuremath{p_T}}
\newcommand{\sqrtsnn}{\ensuremath{\sqrt{s_{_{\!N\!N}}}}}
\newcommand{\measresult}[4]{%
  \ensuremath{#1%
    \ifthenelse{\equal{#2}{}}%
    {}%
    {\pm #2%
      \ifthenelse{\equal{#3}{}}%
      {}%
      {_\text{stat.}}%
    }%
    \ifthenelse{\equal{#3}{}}%
    {}%
    {\pm #3_\text{syst.}}\text{#4}%
  }%
}

\newcommand{\mevc}{~\ensuremath{\text{MeV}\! / c}}

\newcommand{\mevcc}{~\ensuremath{\text{MeV}\! / c^2}}
\newcommand{\gevcc}{~\ensuremath{\text{GeV}\! / c^2}}
\newcommand{\tenpow}[2][]{%
  \ifthenelse{\equal{#1}{}}
  {\ensuremath{10^{#2}}}
  {\ensuremath{{#1} \cdot 10^{#2}}}
}

\newcommand{\figref}[1]{{Fig.~\ref{fig:#1}}}
\newcommand{\Figref}[1]{{Figure~\ref{fig:#1}}}
\newcommand{\equref}[1]{{Eq.~\eqref{eq:#1}}}

\newcommand{\secref}[1]{{section~\ref{sec:#1}}}

\newcommand{\others}{\textit{et al.}}

\begin{document}

%\pagewiselinenumbers

% front matter
\title{Observation of $\bm{\fourpion}$ Photoproduction in
  Ultra-Peripheral Heavy Ion Collisions at $\bm{\sqrtsnn = 200}$~GeV at
  the STAR Detector}

\date{\today}

\affiliation{Argonne National Laboratory, Argonne, Illinois 60439, USA}
\affiliation{University of Birmingham, Birmingham, United Kingdom}
\affiliation{Brookhaven National Laboratory, Upton, New York 11973, USA}
\affiliation{University of California, Berkeley, California 94720, USA}
\affiliation{University of California, Davis, California 95616, USA}
\affiliation{University of California, Los Angeles, California 90095, USA}
\affiliation{Universidade Estadual de Campinas, Sao Paulo, Brazil}
\affiliation{University of Illinois at Chicago, Chicago, Illinois 60607, USA}
\affiliation{Creighton University, Omaha, Nebraska 68178, USA}
\affiliation{Czech Technical University in Prague, FNSPE, Prague, 115 19, Czech Republic}
\affiliation{Nuclear Physics Institute AS CR, 250 68 \v{R}e\v{z}/Prague, Czech Republic}
\affiliation{University of Frankfurt, Frankfurt, Germany}
\affiliation{Institute of Physics, Bhubaneswar 751005, India}
\affiliation{Indian Institute of Technology, Mumbai, India}
\affiliation{Indiana University, Bloomington, Indiana 47408, USA}
\affiliation{University of Jammu, Jammu 180001, India}
\affiliation{Joint Institute for Nuclear Research, Dubna, 141 980, Russia}
\affiliation{Kent State University, Kent, Ohio 44242, USA}
\affiliation{University of Kentucky, Lexington, Kentucky, 40506-0055, USA}
\affiliation{Institute of Modern Physics, Lanzhou, China}
\affiliation{Lawrence Berkeley National Laboratory, Berkeley, California 94720, USA}
\affiliation{Massachusetts Institute of Technology, Cambridge, MA 02139-4307, USA}
\affiliation{Max-Planck-Institut f\"ur Physik, Munich, Germany}
\affiliation{Michigan State University, East Lansing, Michigan 48824, USA}
\affiliation{Moscow Engineering Physics Institute, Moscow Russia}
\affiliation{City College of New York, New York City, New York 10031, USA}
\affiliation{NIKHEF and Utrecht University, Amsterdam, The Netherlands}
\affiliation{Ohio State University, Columbus, Ohio 43210, USA}
\affiliation{Old Dominion University, Norfolk, VA, 23529, USA}
\affiliation{Panjab University, Chandigarh 160014, India}
\affiliation{Pennsylvania State University, University Park, Pennsylvania 16802, USA}
\affiliation{Institute of High Energy Physics, Protvino, Russia}
\affiliation{Purdue University, West Lafayette, Indiana 47907, USA}
\affiliation{Pusan National University, Pusan, Republic of Korea}
\affiliation{University of Rajasthan, Jaipur 302004, India}
\affiliation{Rice University, Houston, Texas 77251, USA}
\affiliation{Universidade de Sao Paulo, Sao Paulo, Brazil}
\affiliation{University of Science \& Technology of China, Hefei 230026, China}
\affiliation{Shandong University, Jinan, Shandong 250100, China}
\affiliation{Shanghai Institute of Applied Physics, Shanghai 201800, China}
\affiliation{SUBATECH, Nantes, France}
\affiliation{Texas A\&M University, College Station, Texas 77843, USA}
\affiliation{University of Texas, Austin, Texas 78712, USA}
\affiliation{Tsinghua University, Beijing 100084, China}
\affiliation{United States Naval Academy, Annapolis, MD 21402, USA}
\affiliation{Valparaiso University, Valparaiso, Indiana 46383, USA}
\affiliation{Variable Energy Cyclotron Centre, Kolkata 700064, India}
\affiliation{Warsaw University of Technology, Warsaw, Poland}
\affiliation{University of Washington, Seattle, Washington 98195, USA}
\affiliation{Wayne State University, Detroit, Michigan 48201, USA}
\affiliation{Institute of Particle Physics, CCNU (HZNU), Wuhan 430079, China}
\affiliation{Yale University, New Haven, Connecticut 06520, USA}
\affiliation{University of Zagreb, Zagreb, HR-10002, Croatia}

\author{B.~I.~Abelev}\affiliation{University of Illinois at Chicago, Chicago, Illinois 60607, USA}
\author{M.~M.~Aggarwal}\affiliation{Panjab University, Chandigarh 160014, India}
\author{Z.~Ahammed}\affiliation{Variable Energy Cyclotron Centre, Kolkata 700064, India}
\author{A.~V.~Alakhverdyants}\affiliation{Joint Institute for Nuclear Research, Dubna, 141 980, Russia}
\author{B.~D.~Anderson}\affiliation{Kent State University, Kent, Ohio 44242, USA}
\author{D.~Arkhipkin}\affiliation{Brookhaven National Laboratory, Upton, New York 11973, USA}
\author{G.~S.~Averichev}\affiliation{Joint Institute for Nuclear Research, Dubna, 141 980, Russia}
\author{J.~Balewski}\affiliation{Massachusetts Institute of Technology, Cambridge, MA 02139-4307, USA}
\author{L.~S.~Barnby}\affiliation{University of Birmingham, Birmingham, United Kingdom}
\author{S.~Baumgart}\affiliation{Yale University, New Haven, Connecticut 06520, USA}
% ...

\collaboration{STAR Collaboration}\noaffiliation

%%% Local Variables: 
%%% mode: latex
%%% TeX-master: "4ProngPaper"
%%% End: 

%\pagewiselinenumbers

\break
\begin{abstract}
  We present a measurement of \fourpion\ photonuclear production in
  ultra-peripheral \PAu-\PAu\ collisions at $\sqrtsnn = 200$~GeV from
  the STAR experiment. The \fourpion\ final states are observed at low
  transverse momentum and are accompanied by mutual nuclear excitation
  of the beam particles. The strong enhancement of the production
  cross section at low transverse momentum is consistent with coherent
  photoproduction. The \fourpion\ invariant mass spectrum of the
  coherent events exhibits a broad peak around
  \measresult{1540}{40}{}{\mevcc} with a width of
  \measresult{570}{60}{}{\mevcc}, in agreement with the photoproduction
  data for the $\Prz(1700)$. We do not observe a corresponding peak in
  the \twopion~final state and measure an upper limit for the ratio of
  the branching fractions of the $\Prz(1700)$ to \twopion\ and
  \fourpion\ of 2.5~\% at 90~\% confidence level. The ratio of
  $\Prz(1700)$ and $\Prz(770)$~coherent production cross sections is
  measured to be \measresult{13.4}{0.8}{4.4}{\%}.
%\vspace{1pc}
\end{abstract}


\maketitle

\section{Introduction}
\label{sec:introduction}
The electromagnetic field of a relativistic heavy nucleus can be
approximated by a flux of quasi-real virtual photons using the
Weizs\"acker-Williams approach~\cite{weizsacker_williams}. Because the
number of photons grows with the square of the nuclear charge, fast
moving heavy ions generate intense photon fluxes. Relativistic heavy
ions can thus be used as photon sources or targets. Due to the long
range of the electromagnetic interactions, they can be separated from
the hadronic interactions by requiring impact parameter~$b$ larger
than the sum of the nuclear radii~$R_\PA$ of the beam particles. These
so-called ultra-peripheral heavy ion collisions~(UPCs) allow us to
study photonuclear effects as well as photon-photon
interactions~\cite{baur_krauss_bertulani_hencken}.

A typical high-energy photonuclear reaction in UPCs is the
production of vector mesons. In this process the virtual photon,
radiated by the ``emitter'' nucleus, fluctuates into a virtual
\Pqq\Pqqbar-pair, which scatters elastically off the ``target''
nucleus, thus producing a real vector meson. The scattering can be
described in terms of soft Pomeron exchange. The cross section for
vector meson production depends on how the virtual \Pqq\Pqqbar-pair
couples to the target nucleus. This is mainly determined by the
transverse momentum~\pT\ of the produced meson. For small transverse
momenta of the order of $\pT \lesssim \hbar / R_\PA$ the
\Pqq\Pqqbar-pair couples coherently to the entire nucleus.  This leads
to large cross sections which depend on the nuclear form factor
$F(t)$, where $t$ is the square of the momentum transfer to the target
nucleus. For larger transverse momenta the \Pqq\Pqqbar-pair couples to
the individual nucleons in the target nucleus. This ``incoherent''
scattering has a smaller cross section that scales approximately with
the mass number~$A$ modulo corrections for nuclear absorption of the
meson.

\begin{figure}[htb]
  \includegraphics[width=0.8\linewidth]{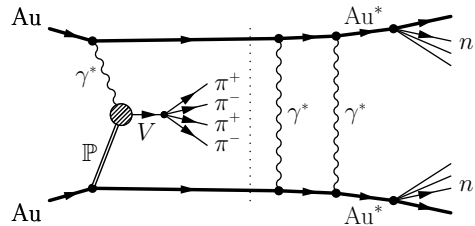}
  \vspace{-3ex}
  \caption{Schematic view of the photonuclear production of a vector
    meson~$V$ in an ultra-peripheral \PAu-\PAu\ collision and its
    subsequent decay into four charged pions.  The meson production in
    the fusion processes of photon \PvPh\ and Pomeron \PPom\ is
    accompanied by mutual Coulomb excitation of the beam ions. The
    processes are independent as indicated by the dotted line.}
  \label{fig:vectorMesonProd}
\end{figure}

Due to the intense photon flux in UPCs, it is possible that vector
meson production is accompanied by Coulomb excitation of the beam
particles. The excited ions mostly decay via the emission of
neutrons~\cite{baltz_baur} which is a distinctive event signature that
is utilized in the trigger decision. To lowest order, events with
mutual nuclear dissociation are described by three-photon exchange
(see \figref{vectorMesonProd}): one photon to produce the vector meson
and two photons to excite the nuclei. All three photon exchanges are
in good approximation independent, so that the cross section for the
production of a vector meson~$V$ accompanied by mutual nuclear
dissociation can be factorized~\cite{baltz_baur}:

\begin{multline}
  \sigma_{V,\, x\Pn\, x\Pn} = \\
    \int\!\D[2]{b} \lrBrk{1 - P_\text{had}(b)} \cdot P_V(b) \cdot P_{\!x\Pn, 1}(b) \cdot P_{\!x\Pn, 2}(b),
  \label{eq:sigma_mutual_excitation}
\end{multline}

\noindent
where $P_\text{had}(b)$~is the probability for hadronic interaction,
$P_V(b)$~the probability to produce a vector meson~$V$\!\!, and
$P_{\!x\Pn, i}(b)$ the probability that nucleus~$i$ emits
$x$~neutrons. Compared to exclusive photonuclear vector meson
production, reactions with mutual Coulomb excitation have smaller
median impact parameters.

The PDG currently lists two excited \Prz~states, the $\Prz(1450)$ and
the $\Prz(1700)$, which are seen in various production modes and decay
channels including two- and four-pion final states~\cite{pdg}. The
nature of these states is still an open question, because
their decay patterns do not match quark model
predictions~\cite{rho_prime}. Little data exist on high-energy
photoproduction of excited \Prz~states in the four-pion decay
channel. Most of them are from photon-proton or
photon-deuteron fixed target experiments at photon energies in the
range from 2.8 to 18~GeV~\cite{four_pion_photo_prod_low_e_1,
  four_pion_photo_prod_low_e_2, four_pion_photo_prod_low_e_3,
  four_pion_photo_prod_low_e_4}. The OMEGA spectrometer measured
photoproduction on proton targets at energies $E_\PPh$ of up to
70~GeV~\cite{four_pion_omega}. The heaviest target nucleus used so far
to study diffractive two- and four-pion photoproduction was carbon
with photon energies between 50 and
200~GeV~\cite{four_pion_fnal}. These experiments observe a broad
structure in the four-pion invariant mass distribution at masses
ranging from
\measresult{1430}{50}{}{\mevcc}~\cite{four_pion_photo_prod_low_e_1} to
\measresult{1570}{60}{}{\mevcc}~\cite{four_pion_photo_prod_low_e_3}
and with widths between
\measresult{340}{60}{}{\mevcc}~\cite{four_pion_photo_prod_low_e_3} and
\measresult{850}{200}{}{\mevcc}~\cite{four_pion_photo_prod_low_e_2} that
the PDG assigns to the $\Prz(1700)$. However, data indicate that the
peak might consist of two
resonances~\cite{four_pion_photo_prod_low_e_4}. We will use the symbol
$\Pr'$ to designate this structure in the rest of the text.

The measurements presented in this paper extend the four-pion
photoproduction data to fixed target equivalent photon energies of up
to 320~GeV as well as to heavy target nuclei. This represents
the first measurement of four-prong production in ultra-peripheral
heavy ion collisions complementing the pioneering work on $\Pep\Pem$,
$\Prz(770)$, and \PJ~production in UPCs at STAR~\cite{star_upc_rho_1,
  star_upc_rho_2, star_upc_rho_int, star_upc_ee} and
PHENIX~\cite{phenix_upc_jpsi}.

There are at least three models for the production of $\Prz(770)$
mesons in ultra-peripheral collisions: The model of Klein and
Nystrand~(KN)~\cite{KN} employs the Vector Dominance Model~(VDM) to
describe the virtual photon and a classical mechanical approach for
the scattering on the target nucleus, using results from $\PPh\, \Pp
\to \Prz(770)\, \Pp$ experiments. The Frankfurt, Strikman, and
Zhalov~(FSZ) model~\cite{FSZ} is based on a generalized VDM for the
virtual photon and a QCD Gribov-Glauber approach for the
scattering. The model of Gon\c{c}alves and Machado~(GM)~\cite{GM}
employs a QCD color dipole approach that takes into account nuclear
effects and parton saturation phenomena. The KN~model agrees best with
the available data on $\Prz(770)$ production, the FSZ and in
particular the GM model overestimate the $\Prz(770)$ production cross
section~\cite{star_upc_rho_2}.  Only the FSZ calculations make
predictions about the production of exited \Prz~states in UPCs.

\section{Experimental Setup and Data Selection}
\label{sec:data_sel}
The analysis is based on \tenpow[1.9]{6}~events taken with the STAR
experiment at the Relativistic Heavy Ion Collider~(RHIC) in \PAu-\PAu\
collisions at $\sqrtsnn = 200$~GeV during the year 2007 run. The
Solenoidal Tracker at RHIC (STAR) experiment uses a large cylindrical
Time Projection Chamber~(TPC)~\cite{star_tpc} with 2~m radius and
4.2~m length, operated in a 0.5~T solenoidal magnetic field, to
reconstruct charged tracks.

Two detector systems are used for triggering: the Central Trigger
Barrel~(CTB)~\cite{star_ctb}, which is an array of 240~plastic
scintillator slats around the TPC that allows us to trigger on charged
particle multiplicities, and the two Zero Degree
Calorimeters~(ZDCs)~\cite{star_zdc}, which are located at $\pm 18$~m
from the interaction point. The ZDCs have an acceptance close to unity
for neutrons originating from nuclear dissociation of the beam
ions. In the trigger, these neutrons are used to tag UPC events by
requiring coincident hits in both ZDCs with amplitudes corresponding
to less than about~7 to 10~neutrons. The ADC sum of all CTB slats is
restricted to a range equivalent to a hit multiplicity between about~2
and~40 minimum ionizing particles. In order to enrich events at
central rapidities, events with hits above the minimum ionizing
particle threshold in the large-tile Beam-Beam Counters
(BBCs)~\cite{star_bbc}, which cover $2.1 < |\eta| < 3.6$, are vetoed.

\begin{figure}[htb]
  \includegraphics[width=1.03\linewidth]{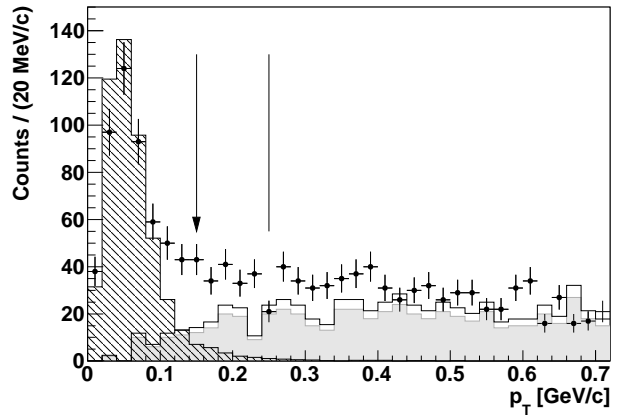}
  \vspace{-3ex}
  \caption{Distribution of the \fourpion\ transverse momentum $\pT =
    \lrabs{\sum_{i = 1}^4 \vec{p}_{T,i}}$: The filled circles are the
    measured points with the statistical errors. The hatched filled
    histogram shows the expected distribution from simulation of
    coherent photoproduction (cf. \secref{acc_corr}). The strong
    enhancement at low transverse momenta is due to coherently
    produced \fourpion. This unique signature is used in the event
    selection which requires $\pT < 150\mevc$ (arrow). The remaining
    background is estimated from $+2$ or $-2$ charged four-prong
    combinations, by normalizing ($\text{factor} = 1.186 \pm 0.054$)
    their \pT\ distribution (gray filled histogram) to that of the
    neutral four-prongs in the region of $\pT > 250\mevc$ (vertical
    line) yielding the unfilled histogram (see \secref{results}).}
  \label{fig:rhoPrimePtFull}
\end{figure}

In the offline analysis two- and four-prong data sets are selected.
Four-prong events are required to have exactly four tracks with zero
net charge in the TPC that form a common (primary) vertex. Because the
STAR TPC has a drift time of about 36~\textmu sec, any charged tracks
produced within a time window of $\pm 36$~\textmu sec around the
triggered collision will overlap with the event of interest. Some of
these additional tracks come from beam induced background reactions,
but, due to the high luminosities reached in the RHIC 2007
run~\cite{rhic_run7}, a large percentage is from real heavy ion
collisions. In order to account for those out-of-time events and
backgrounds up to 86~additional tracks per event, which do not point
to the primary vertex, are allowed, but excluded from the analysis.
The primary vertex is confined to a cylindrical region of 15~cm radius
and 200~cm length centered around the interaction diamond which
reduces contaminations from pile-up events and beam-gas
interactions. Each of the four-prong tracks is required to have at
least~14 out of a maximum possible 45~hits in the TPC. No particle
identification is employed in the event selection; all four
tracks are assumed to be pions. The distribution of the
  ionization energy loss $\D{E}\! /\! \D{x}$ of the selected tracks in the
  TPC indicates that contaminations from other particle species are
  small. The transverse momentum distribution of the \fourpion\
combinations, as shown in \figref{rhoPrimePtFull}, exhibits an
enhancement at low~\pT, characteristic of coherent
production. Coherent events are selected by requiring $\pT <
150\mevc$. This cut also suppresses contaminations from peripheral
hadronic interactions and from $\fourpion + X$ events, where the $X$
is not reconstructed.

Due to charge conjugation invariance, we expect no $\Prz(770)\,
\Prz(770)$ component in the diffractively produced \fourpion\ final
state. Possible contributions from $\Prz(770)$ pair production by two
independent photoproduction reactions on the same ion pair are
negligible. The KN~model predicts a cross section ratio of exclusive
photonuclear \Prz~pair production and exclusive single \Prz~production
of about $1.2 \cdot 10^{-3}$~\cite{KN}. For mutual nuclear
dissociation of the beam ions the ratio is expected to be of
comparable value so that contaminations of the \fourpion\ sample by
this process are at most a few percent. Also $\PvPh \PvPh \to
\Prz(770)\, \Prz(770)$ events contribute below the percent level. Here
the cross section ratio for exclusive \Prz~pair production in
two-photon events and exclusive photonuclear \Prz(770)~production was
calculated to be $3.2 \cdot 10^{-5}$ for $\Prz(770)$ pair invariant
masses in the range between 1.5 and
1.6\gevcc~\cite{rho_pair_two_photon}.

The two-prong selection criteria are very similar and follow the STAR
UPC $\Prz(770)$ analyses~\cite{star_upc_rho_1, star_upc_rho_2}. As in
the four-prong case, out-of-time events and background are taken into
account by allowing up to 36~tracks per event in addition to the two
primary TPC tracks.  Background from two-photon \Pep\Pem\ and
photonuclear \Po~production is
negligible~\cite{star_upc_rho_2}. Cosmic ray background is strongly
suppressed, due to the ZDC requirement in the trigger.

\section{Efficiency and Acceptance Corrections}
\label{sec:acc_corr}
Detector efficiency and acceptance are studied using a Monte Carlo
event generator based on the KN~model~\cite{KN} which describes
coherent vector meson production accompanied by mutual Coulomb
excitation in UPCs. In order to reduce model dependence, the
acceptance corrections are applied in two stages.  Within the detector
acceptance of $\abs{y} < 1$, the corrections are calculated using a
realistic detector simulation based on GEANT~3~\cite{geant}. In a
second step, the results are then extrapolated to the full $4 \pi$
solid angle based on the KN~model distributions.

In order to determine the acceptance corrections for the four-prong
case, we assume a simple decay model, where an excited \Prz~meson
decays into $\Prz(770)$ and $\Pf(600)$, each in turn decaying into
\twopion:

\begin{equation}
  \Pr' \to \Prz(770)\,\, \Pf(600) \to [\twopion]_\text{$P$-wave}\,\, [\twopion]_\text{$S$-wave}
  \label{eq:decay_model}
\end{equation}

\begin{figure}[t]
  \includegraphics[width=1.03\linewidth]{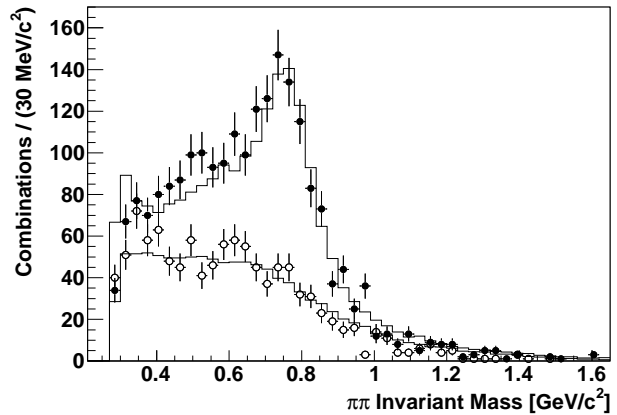}
  \vspace{-3ex}
  \caption{Invariant Mass distribution of two-pion subsystems: The
    filled circles show the measured \twopion\ invariant mass spectrum
    for the selected four-prong sample (four entries per event) with
    statistical errors. The open circles represent the mass spectrum
    of the like-sign pion pairs (two entries per event). The
    unlike-sign mass distribution exhibits an enhancement with respect
    to the like-sign pairs in the $\Prz(770)$ region. The solid line
    histograms show the prediction from simulation assuming the
    relative $S$-wave decay $\Pr' \to \Prz(770)\,\, \Pf(600)$.}
  \label{fig:rhoPrimeMass2Pi}
\end{figure}

\begin{figure}[h]
  \includegraphics[width=1.03\linewidth]{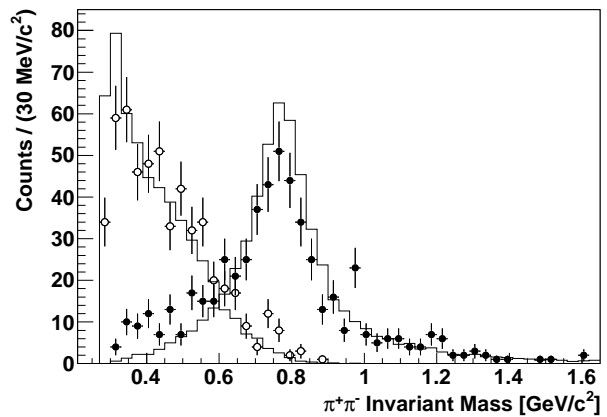}
  \vspace{-3ex}
  \caption{Invariant Mass distribution of two-pion subsystems: The
    open circles show the measured invariant mass spectrum of the
    lightest \twopion\ pair in the event with the bars indicating the
    statistical errors.  The filled circles represent the invariant
    mass distribution of the \twopion\ that is recoiling against the
    lightest pair. The spectrum exhibits a clear peak in the
    $\Prz(770)$ region. The solid line histograms show the prediction
    from simulation assuming the relative $S$-wave decay $\Pr' \to
    \Prz(770)\,\,\Pf(600)$.}
  \label{fig:rhoPrimeMass2PiMin}
\end{figure}

This decay model is motivated by the fact that the invariant mass
spectrum of the unlike-sign two-pion subsystems in the four-prong
sample shows an enhancement around the $\Prz(770)$ mass
(cf. \figref{rhoPrimeMass2Pi}). \Figref{rhoPrimeMass2PiMin} compares
the invariant mass spectrum of the lightest \twopion\ pair with the
spectrum of the pair recoiling against it and shows that the four-pion
final state consists mainly of a low-mass pion pair accompanied by a
$\Prz(770)$.

In principle, the \Prz\ and \Pf\ are allowed to be in a relative $S$-
or $D$-wave, but, due to the low statistics of the data, we are not
able to estimate the $D$-wave parameters. Consequently we only
consider $S$-wave decay. Possible $D$-wave contributions are well
within the estimated systematic error (see \secref{results}).

 The angular distribution~$I$ that is used to estimate the acceptance
corrections for the four-prong sample is parameterized:

\begin{equation}
  I \propto \sum_\epsilon \sum_{m, m'} {^{\epsilon}r_{m, m'}} \cdot {^\epsilon\!\!\mathcal{A}^J_{m}}
            \cdot {^\epsilon\!\!\mathcal{A}^{J*}_{m'}},
  \label{eq:ang_distr}
\end{equation}
where $^\epsilon\!\!\mathcal{A}^J_{m}$ is the amplitude for the decay
of a $\Pr'$ with spin $J = 1$ and a projection $m$ of $J$ along the
quantization axis assuming the model of \equref{decay_model}.
$^{\epsilon}r_{m, m'}$ represents the spin density matrix elements.
The amplitudes are defined in the $\Pr'$ rest frame with the $z$-axis
along the beam direction and the $y$-axis parallel to the production
plane normal, $\vec{p}_\text{beam} \times \vec{p}_{\Pr'}$.  Due to the
large beam energy and the coherent nature of the production process,
this frame coincides approximately with the $\Pr'$ helicity frame.
Both the amplitudes and the spin density matrix are constructed using
eigenstates of the operator $\Pi_y$ of reflections in the production
plane, the so-called reflectivity basis with $\epsilon = \pm
1$~\cite{reflectivity}.  The sum in \equref{ang_distr} is simplified
by assuming $s$-channel helicity conservation~(SCHC) and that the
quasi-real photons come with helicities $\pm 1$ only, so that
${^-\!r_{11}} = {^+\!r_{11}}$ are the only non-zero spin density
matrix elements. The amplitudes $^\epsilon\!\!\mathcal{A}^J_{m}$ are
factorized:

\begin{equation}
 {^\epsilon\!\!\mathcal{A}^J_{m}} = \Delta_\Pr(m_\Pr) \cdot \Delta_{\Pf}(m_{\Pf}) \cdot 
                                  {^\epsilon\!\mathcal{M}^J_{m}}(\theta, \phi; \theta_\Pr, \phi_\Pr, \gamma_\Pr)
  \label{eq:decay_amp}
\end{equation}
Here $\Delta_{\Pr, f_0}$ are the amplitudes for the \Prz\ and \Pf\
resonance shapes as a function of the invariant masses of the
intermediate states $m_{\Pr, f_0}$. For the $\Prz(770)$ a $P$-wave
Breit-Wigner with mass-dependent width including Blatt-Weisskopf
barrier factors is used~\cite{blatt_weisskopf}. The \Pf\ is modeled by
an $S$-wave Breit-Wigner at 400\mevcc\ with a width of 600 \mevcc. The
decay amplitudes ${^\epsilon\!\mathcal{M}^J_{m}}$ describe the angular
dependence and include relativistic corrections via the Lorentz factor
$\gamma_\Pr$ of the \Prz\ in the $\Pr'$ rest frame~(RF) according
to~\cite{decay_amp}. ${^\epsilon\!\mathcal{M}^J_{m}}$ depends on the
angles $\theta$ and $\phi$ of the \Prz\ in the $\Pr'$ rest frame as
well as on the angles $\theta_\Pr$ and $\phi_\Pr$ that describe the
orientation of the~\Ppip\ from the \Prz~decay in the \Prz~helicity
rest frame. This frame is defined starting from the $\Pr'$ rest frame
and has its $z_h$-axis parallel to the \Prz-momentum and its
$y_h$-axis along the cross product of beam and \Prz~momentum.
Finally, the sums in \equref{ang_distr} are Bose symmetrized to
account for the four indistinguishable final state configurations.

The simulations agree well with the two- and four-pion kinematic
distributions. The mean \Prz~reconstruction efficiency in the region
$\abs{y} < 1$ is about \measresult{21.9}{0.2}{}{\%}, that for
the~$\Pr'$ approximately \measresult{6.5}{0.5}{}{\%}. The efficiencies
show no strong dependence on the $z$-position of the primary vertex or
on the transverse momentum in the region of the coherent
peak. However, due to the TPC acceptance, the efficiencies decrease to
roughly 1~\% for the~\Prz\ and 0.1~\% for the~$\Pr'$, respectively, at
$y = \pm 1$. The mass dependence of the \Prz~efficiency is flat for
masses above about 600\mevcc\ and decreases quickly for lower
masses. The $\Pr'$~efficiency rises with mass, until it reaches a
plateau at approximately 1500\mevcc, so that the \fourpion\ mass peak
in the uncorrected data is shifted towards larger masses (see the
dashed curve in \figref{rhoPrimeMass}).

From the simulations the resolutions for \pT, $y$, and invariant
mass of the selected pion pairs are estimated to be
approximately 6\mevc, 0.009, and 5\mevcc, respectively. The
corresponding values for the four-pion combinations are 10\mevc,
0.006, and 10\mevcc.

\section{Results}
\label{sec:results}
The ratio of coherent $\Pr'$ and $\Prz(770)$ production cross sections
can be calculated from the respective acceptance-corrected yields
which are determined from fits of the \fourpion\ and \twopion\
invariant mass distributions, respectively.

\begin{figure}[t]
  \hspace*{-0.5em}
  \includegraphics[width=1.03\linewidth]{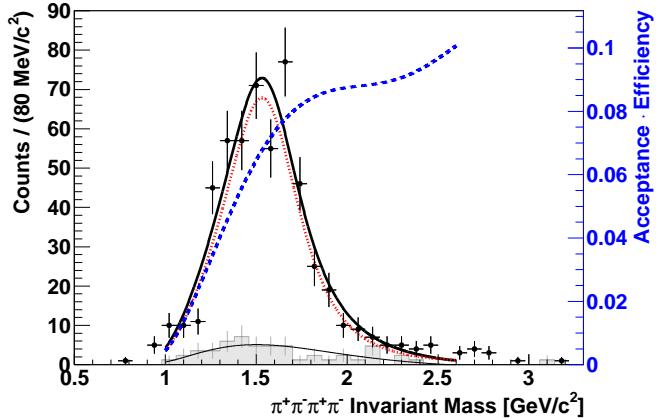}
  \vspace{-3ex}
  \caption{(Color online) Invariant mass distribution of coherently
    produced \fourpion: The filled circles are the measured points
    with the statistical errors, the gray filled histogram is the
    background estimated from charged four-prongs (cf.
    \figref{rhoPrimePtFull}). The thick black line shows the fit of a
    modified $S$-wave Breit-Wigner on top of a second order polynomial
    background (thin black line; cf. \equref{rhoPrimeMass}) taking
    into account the detector acceptance in the region $\abs{y} < 1$
    (rising dashed line). The dotted curve represents the signal curve
    without background.}
  \label{fig:rhoPrimeMass}
\end{figure}

\Figref{rhoPrimeMass} shows the measured \fourpion\ invariant mass
spectrum which exhibits a broad peak around 1540\mevcc\ indicating
resonant $\Pr'$ production similar to what was seen in fixed-target
photoproduction experiments~\cite{four_pion_photo_prod_low_e_1,
  four_pion_photo_prod_low_e_2, four_pion_photo_prod_low_e_3,
  four_pion_photo_prod_low_e_4, four_pion_omega, four_pion_fnal}. This
assumes that the peak is dominated by spin states with quantum numbers
$J^{PC} = 1^{--}$. Contributions from other spin states cannot be
ruled out, because in order to disentangle them a much larger data set
would be required.

The data are fit in the range from~1 to 2.6\gevcc\ with a relativistic
$S$-wave Breit-Wigner which is modified by the phenomenological
Ross-Stodolsky factor~\cite{ross_stodolsky} and which sits on top of a
second order polynomial that parameterizes the remaining combinatorial
background:

\begin{equation}
  f_{4 \Ppi}(m) = A \cdot \lrbrk{\frac{m_0}{m}}^n\!\!\cdot \frac{m_0^2
    \Gamma_0^2}{(m_0^2 - m^2)^2 + m_0^2 \Gamma_0^2} + f_\text{BG}(m)
  \label{eq:rhoPrimeMass}
\end{equation}
Here $m$ is the \fourpion\ invariant mass.  The resonance mass $m_0$,
the width $\Gamma_0$, and the exponent $n$ are left as free
parameters.

The background polynomial $f_\text{BG}$ is fixed by fitting the
invariant mass distribution of $+2$ or $-2$ charged four-prongs.
Because at larger \pT\ the coherent cross section becomes negligible,
the region $\pT > 250\mevc$ is used to define the total amount of
background by scaling the \pT~distribution of the charged four-prongs,
so that it matches that of the neutral four-prongs
(cf. \figref{rhoPrimePtFull}). This procedure treats incoherently
produced \fourpion\ as background. The extracted scaling factor
of \measresult{1.186}{0.054}{}{} --- about half of the value estimated
for the \Prz\ background (see below) --- is applied to the background
polynomial.

Fitting \equref{rhoPrimeMass} to the data yields a resonance mass of
\measresult{1540}{40}{}{\mevcc}, a width of
\measresult{570}{60}{}{\mevcc}, and an exponent of $n =
\measresult{2.4}{0.7}{}{}$. The values for mass and width, however,
depend strongly on the choice of $n$. The peak contains $N_{4 \Ppi} =
\measresult{9180}{540}{}{~events}$ in the mass range from~1
to~2.6\gevcc. As seen in \figref{rhoPrimeMass} and also indicated by
the $\chi^2 / \text{n.d.f.}$ of the maximum likelihood fit of about
$36/16$, \equref{rhoPrimeMass} does not describe the peak shape
well. This is in accord with observations from other photo-production
experiments, which favor a description using two resonances in this
mass region~\cite{four_pion_photo_prod_low_e_4}. However, the low
  statistics of the data does not permit to extract the resonance
  and mixing parameters for a two-resonance scenario.

In both the background and the signal fit, the mass dependence of the
reconstruction efficiency for $\abs{y} < 1$ is taken into account
(dashed curve in \figref{rhoPrimeMass}).  The efficiency is
parameterized by a fifth order polynomial determined by fitting the
Monte Carlo data.

\begin{figure}[t]
  \includegraphics[width=1.03\linewidth]{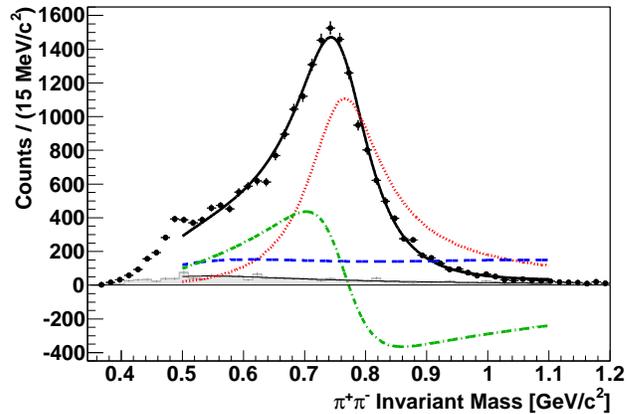}
  \vspace{-3ex}
  \caption{(Color online) Invariant mass distribution of coherently
    produced \twopion\ pairs. The filled circles are the measured
    points with statistical errors. The thick black line shows the fit
    taking into account the detector acceptance in the region $\abs{y}
    < 1$. The non-interfering combinatorial background is represented
    by the thin black line, which is a fit to the like-sign invariant
    mass distribution scaled by a factor estimated from the \pT\
    distribution (gray filled histogram). The dotted curve shows the
    Breit-Wigner without background, the dashed line the interfering
    background component that is assumed to be mass-independent. The
    dash-dotted curve is the S\"oding interference term of the
    two~\cite{soeding_term}.}
  \label{fig:rhoMass}
\end{figure}

The $\Prz(770)$~peak in the \twopion\ invariant mass distribution of
the selected two-prong data set is fit by a $P$-wave Breit-Wigner with
mass-dependent width and S\"oding interference
term~\cite{soeding_term} on top of a second order polynomial
background as described in~\cite{star_upc_rho_1, star_upc_rho_2,
  star_upc_rho_int} (cf. \figref{rhoMass}). As in the $\Pr'$ case, the
background polynomial is determined from a fit of the like-sign pair
invariant mass distribution that is scaled up by a factor of
\measresult{2.284}{0.050}{}{} which is extracted from the incoherent
part of the \pT\ distribution. The fit gives a \Prz~mass of
\measresult{772.3}{1.2}{}{\mevcc} and a width of
\measresult{152.1}{1.9}{}{\mevcc}, in agreement with both the PDG data
on \Prz\ photoproduction~\cite{pdg} and earlier results from
photonuclear production~\cite{star_upc_rho_1, star_upc_rho_2,
  star_upc_rho_int}. As expected, modifications of the $\Prz(770)$
properties that were measured in peripheral \PAu-\PAu\
collisions~\cite{star_rho_in_medium} and attributed to in-medium
production are not observed in the current study. The Breit-Wigner
peak contains $N_\Pr = \measresult{55\,940}{910}{}{~events}$ in the
mass range from~500 to 1100~\mevcc. The $\chi^2/\text{n.d.f.}$ of the
maximum likelihood fit is $115 / 36$ which mainly reflects the fact
that the fit function does not reproduce well either the high mass
tail of the $\Pr(770)$ or the low mass region.  This mass region
exhibits a peak from $K_s^0 \to \twopion$, where the kaons most likely
come from photoproduced $\phi(1020)$.

Using the acceptance-corrected yields $N_\Pr$ and $N_{4 \Ppi}$ for the
$\Prz(770)$ and the $\Pr'$, respectively, it is possible to calculate
the cross section ratio for coherent \Prz\ and $\Pr'$ production which
is accompanied by mutual nuclear excitation and where the $\Pr'$
decays into \fourpion:

\begin{equation}
  \frac{\sigma_{4 \Ppi, x\Pn\,x\Pn}^\text{coh} }{\sigma_{\Pr,\, x\Pn\,x\Pn}^\text{coh}}
  = \frac{\sigma_{\Pr'\!\!,\, x\Pn\,x\Pn}^\text{coh} \cdot \mathcal{B}(\Pr' \to \fourpion)}{\sigma_{\Pr,\, x\Pn\,x\Pn}^\text{coh}}
  = \frac{N_{4 \Ppi}}{N_\Pr},
\end{equation}
where $\mathcal{B}(\Pr' \to \fourpion)$ is the branching fraction of
the $\Pr'$ into \fourpion.

The cross section ratio does not depend strongly on rapidity and in
the region $\abs{y} < 1$ has a mean value of
\measresult{16.4}{1.0}{5.2}{\%}. Based on the KN~model~\cite{KN} we
estimate extrapolation factors to the full $4 \pi$ solid angle of
\measresult{1.8}{}{0.1}{} for the $\Pr'$ and of
\measresult{2.2}{}{0.1}{} for the \Prz, where the latter value is
taken from~\cite{star_upc_rho_2}.  With this extrapolation, the
overall coherent cross section ratio is
\measresult{13.4}{0.8}{4.4}{\%}. Using the measured cross section
$\sigma_{\Pr,\, x\Pn\,x\Pn}^\text{coh}$ for coherent
$\Prz(770)$~production accompanied by mutual nuclear excitation of the
beam particles from~\cite{star_upc_rho_2}, the $\Pr' \to \fourpion$
production cross section can be calculated. The cross section within
$\abs{y} < 1$ is $\sigma_{4 \Ppi,\, x\Pn\,x\Pn}^\text{coh}(\abs{y} <
1) = \measresult{2.4}{0.2}{0.8}{mb}$, the corresponding
rapidity-integrated value is $\sigma_{4 \Ppi,\, x\Pn\,x\Pn}^\text{coh}
= \measresult{4.3}{0.3}{1.5}{mb}$.

The influence of systematic effects on the cross section ratio was
studied. The main source of systematic uncertainty comes from the
model dependence of the angular distribution of the \fourpion\ used in
the acceptance correction. This uncertainty is estimated to be 21~\%
by comparing to the cross section ratio obtained using an
isotropic angular distribution in the Monte Carlo simulation.
The uncertainty from the parameterization of the $\twopion$ $S$-wave
in the four prong-decay model is about 11~\% and is estimated by
increasing the mass and/or width of the $\Pf(600)$ Breit-Wigner to
600\mevcc\ and 1000\mevcc, respectively. Additional systematic errors
come from the event selection cuts (14~\%), the background subtraction
(10~\%), as well as the invariant mass binning and the fit range
(8~\%). The systematic error associated with the particular choice of
the fit function for the \fourpion\ invariant mass peak
(cf. \equref{rhoPrimeMass}) is estimated to be 9~\% by trying to fit a
non-relativistic Breit-Wigner and by fixing the value of the
Ross-Stodolsky exponent in \equref{rhoPrimeMass} to $n = 0$ and $4$.
The error for the extrapolation to the full $4 \pi$ solid angle was
estimated to be 6~\% for the \Prz\ in~\cite{star_upc_rho_2} by
comparing the KN~\cite{KN} and the FSZ~\cite{FSZ} models. The
extrapolation factor depends on the photon-energy spectrum, which is
well-known, and on the poorly known energy dependence of the
photo-production cross section. Because the KN model~\cite{KN}
describes the observed \fourpion\ rapidity distribution well, we
assume that the $\Pr'$ production mechanism is not too different from
that of the \Prz\ and assign the same systematic error of 6~\%.

The measured cross section ratio cannot be compared directly to the
ratio of the total exclusive coherent $\Pr'$ and \Prz\ cross sections
of 14.2~\% predicted by the FSZ model~\cite{FSZ}, because the
branching fraction for $\Pr' \to \fourpion$ is not known. The ratio
between the cross section for $\Pr'$ production accompanied by mutual
Coulomb excitation, as measured here, and the exclusive coherent
$\Pr'$ cross section, where the beam ions remain unchanged, should be
similar to the one for the \Prz. If in addition a 100~\% branching
fraction to the \fourpion\ final state is assumed, the measured cross
section ratio agrees with the FSZ prediction.  Under the same
assumptions we can estimate, using the value for $\sigma_{\Pr,\,
  0n\,0n}^\text{coh}$ from~\cite{star_upc_rho_2} and the measured
cross section ratio, the total exclusive coherent $\Pr'$ production
cross section to be $\sigma_{\Pr'\!\!,\, 0n\,0n}^\text{coh} =
\measresult{53}{4}{19}{mb}$. The value corresponding to the predicted
cross section ratio is \measresult{56}{3}{8}{mb}. These values are
about half of the exclusive coherent $\Pr'$ cross section of 133~mb
predicted by the FSZ model. On the other hand, this model predicts
also \Prz\ cross section values roughly twice larger than observed by
experiment~\cite{star_upc_rho_2}.

In previous photoproduction experiments using carbon targets the
$\Pr'$ was seen not only in the \fourpion\ decay mode, but also in
\twopion\ final states~\cite{four_pion_fnal}. We do not observe a
significant $\Pr'$ signal in the high mass region of the
$m_{\twopion}$ spectrum as shown in \figref{rhoMassZoom}. In order to
suppress backgrounds, in particular cosmic rays, tighter cuts are
applied: the rapidity is limited to $0.05 < \abs{y} < 1$, the
transverse momentum of the \twopion\ pairs is required to be lower
than 100\mevc, and the primary vertex is confined to
$\abs{z_\text{prim}} < 70$~cm and $r_\text{prim} < 8$~cm.

The $\Pr'$ yield in the resulting \twopion\ invariant mass spectrum is
estimated by fitting the modified $S$-wave Breit-Wigner of
\equref{rhoPrimeMass} on top of an $S$-wave Breit-Wigner for the
high-mass tail of the $\Prz(770)$ in the mass range from 1.1 to
3~\gevcc.  Assuming that the $\Pr'$ peak shape is the same in the
\twopion\ channel, we fixed mass, width, and exponent of the $\Pr'$
Breit-Wigner to the values obtained from the fit of the \fourpion\
invariant mass distribution. This gives an acceptance- and
background-corrected $\Pr'$ yield of $N_{2 \Ppi} =
\measresult{110}{90}{}{}$ in the mass range from 1 to 2.6~\gevcc.
$N_{2 \Ppi}$ can be compared directly to the $\Pr'$ yield $N_{4 \Ppi}$
in the \fourpion\ channel so that the ratio of the branching fractions
of the $\Pr'$ to \twopion\ and to \fourpion\ can be calculated:

\begin{figure}[t]
  \includegraphics[width=1.03\linewidth]{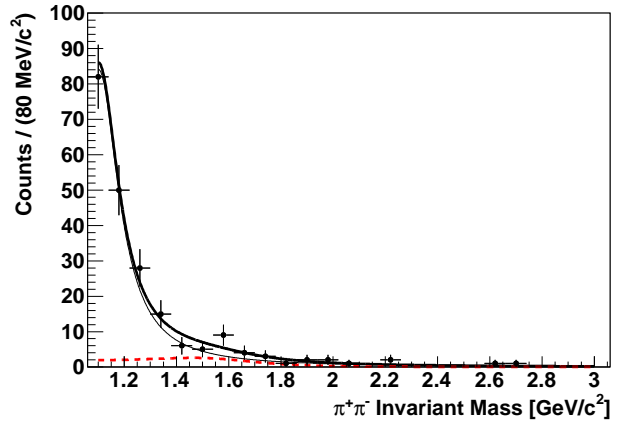}
  \vspace{-3ex}
  \caption{(Color online) High mass region of the $m_{\twopion}$
    spectrum with tighter cuts applied in order to suppress
    background: The filled circles are the measured values with
    statistical errors.  No significant enhancement is seen in the
    region around 1540\mevcc\ where the \fourpion\ invariant mass
    spectrum exhibits a peak. The thick solid line shows the fit of a
    modified $S$-wave Breit-Wigner (cf. \equref{rhoPrimeMass}) with
    parameters fixed to the values extracted from the fit of the
    \fourpion\ invariant mass distribution on top of an $S$-wave
    Breit-Wigner that describes the tail of the $\Prz(770)$ (thin
    solid line) taking into account the detector acceptance. The
    dashed curve represents the signal curve without the \Prz\ tail.}
  \label{fig:rhoMassZoom}
\end{figure}

\begin{equation}
  R = \frac{\mathcal{B(\Pr' \to \twopion)}}{\mathcal{B(\Pr' \to \fourpion)}} = \frac{N_{2 \Ppi}}{N_{4 \Ppi}}
  \label{eq:brRatio}
\end{equation}
Due to the low statistics, the measured value of $R =
\measresult{0.012}{0.010}{}{}$ has a large uncertainty.  The
systematic error from neglecting the $P$-wave nature of the \twopion\
decay by using a mass-independent resonance width in
\equref{rhoPrimeMass} is within the range of the statistical error.
The corresponding upper limit of the ratio is $R < 2.5~\%$ at 90~\%
confidence level.  This is an indication that, in the process measured
here, $R$ is smaller than the ratio of the total $\Pr'$ cross sections
in the two- and four-pion channel on a carbon target which was
measured to be \measresult{6.6}{3.4}{}{\%}~\cite{four_pion_fnal}.

\section{Conclusions}
\label{sec:conlusions}
We have observed diffractive photonuclear production of \fourpion\
final states in ultra-peripheral relativistic heavy ion collisions
accompanied by mutual Coulomb excitation of the beam particles. The
\fourpion\ invariant mass peak exhibits a broad peak around
1540\mevcc. Under the assumption that the peak is dominated by
  spin states with $J^{PC} = 1^{--}$ this is consistent with the
existing photoproduction data currently assigned to the
  $\Prz(1700)$ by the PDG. No corresponding
enhancement in the \twopion\ invariant mass spectrum is found. The
ratio of the branching fractions of the excited \Prz~state to
\twopion\ and \fourpion\ final states is smaller than 2.5~\% at 90~\%
confidence level.  The coherent $\Pr'$ production cross section is
\measresult{13.4}{0.8}{4.4}{\%} of that of the $\Prz(770)$ meson.

\begin{acknowledgments}
  We thank the RHIC Operations Group and RCF at BNL, the NERSC Center at
LBNL and the Open Science Grid consortium for providing resources and
support. This work was supported in part by the Offices of NP and HEP
within the U.S. DOE Office of Science, the U.S. NSF, the Sloan
Foundation, the DFG cluster of excellence `Origin and Structure of the
Universe', CNRS/IN2P3, STFC and EPSRC of the United Kingdom, FAPESP
CNPq of Brazil, Ministry of Ed. and Sci. of the Russian Federation,
NNSFC, CAS, MoST, and MoE of China, GA and MSMT of the Czech Republic,
FOM and NWO of the Netherlands, DAE, DST, and CSIR of India, Polish
Ministry of Sci. and Higher Ed., Korea Research Foundation, Ministry
of Sci., Ed. and Sports of the Rep. Of Croatia, Russian Ministry of
Sci. and Tech, and RosAtom of Russia.

%%% Local Variables: 
%%% mode: latex
%%% TeX-master: "4ProngPaper"
%%% End: 

\end{acknowledgments}

\end{document}